# Post-Quantum Cryptography: $S_{381}$ Cyclic Subgroup of High Order

P. Hecht[1]

*Abstract*—Currently there is an active Post-Quantum Cryptography (PQC) solutions search, which attempts to find cryptographic protocols resistant to attacks by means of for instance Shor's polynomial time algorithm for numerical field problems like integer factorization (IFP) or the discrete logarithm (DLP). The use of non-commutative or non-associative structures are, among others, valid choices for these kinds of protocols. In our case, we focus on a permutation subgroup of high order and belonging to the symmetric group $S_{381}$. Using adequate one-way functions (OWF), we derived a Diffie-Hellman key exchange and an ElGamal ciphering procedure that only relies on combinatorial operations. Both OWF pose hard search problems which are assumed as not belonging to BQP time-complexity class. Obvious advantages of present protocols are their conceptual simplicity, fast throughput implementations, high cryptanalytic security and no need for arithmetic operations and therefore extended precision libraries. Such features make them suitable for low performance and low power consumption platforms like smart cards, USB-keys and cellphones.

*Keywords* – Post-Quantum Cryptography, Non-Commutative Cryptography, Symmetric groups, Permutations, Diffie-Hellman key exchange, ElGamal cipher, Combinatorial analysis.

## 1. INTRODUCTION

Post-Quantum Cryptography (PQC) is a relatively new cryptologic trend [1, 2] that acquired a NIST status [3, 4] and which aims to be resistant to quantum computers attacks (like Shor algorithm [5]). Two main lines of research are non-commutative cryptography (NCC) [6, 7, 8, 9, 10, 11, 12, 13] and non-associative cryptography (NAC) [13, 14, 15, 16, 17]. Belonging to the first category, this paper pursues the development of a fast and cryptanalytically secure solution using high order permutations [18, 19, 20, 21, 22, 23]. The protocol is extremely simple and could be directly adapted to any kind of asymmetric solutions like key exchange, key transport, generalized ElGamal ciphering and ZKP authentication [24, 25, 26, 27, 28, 29, 30, 31, 32, 33]. The keystone here is to work with a high multiplicative order random permutation group <p>, belonging to the non-commutative symmetric group $S_{381}$ [18, 19, 20]. To achieve such performance, a carefully mix of randomness and structured symmetry was designed into the target permutation *p*.

Security of an asymmetric cipher protocol always relies on a one-way function (OWF) [24]. For instance, using the decomposition problem (DP) or the double coset problem (DCP) [7], both assumed to belong to AWPP time-complexity (but out of BQP) [34] problems, which lead to an eventual brute-force attack, thus yielding high computational security.

The cryptographic use of combinatorial structures like permutations is a long-known matter, either in linear way [20] or in two-dimensional combination like Row Latin Squares (RLS) [21, 22] or simply using quasigroups [23]. There are also patented protocols about [35]. Multidimensional tensor solutions are also conceivable, but their utility remains unclear. Other approaches into the same direction are the use of multiple orthogonal latin squares (MOLS) [36] and the use of non-group based latin squares [38]. More information about PQC, NCC and NAC could be found at published works and their own references.

## 2. SOME STRUCTURAL DETAILS

Permutations are simple combinatorial structures [20, 36]. A convenient way to map them as integers is the use of Lehmer's factoradic representation [38, 39]. An optimal random permutation generation with an *O(n)* algorithm is described in [20] as Fisher-Yates-Durstenfeld Algorithm P.

It is a known fact that the order of any permutation is the least common multiple of it independent cycle lengths [40]. So a simple way to construct a random high order group, is to embed any random permutation (say *p*) into prime length cycles using the increasing prime sequence [41] in random order. Summing those cycle lengths; one obtains the symmetric group orders into which the random permutation works as a generator of a cyclic subgroup, whose order is given by the respective primorial function [42]. A valid choice for the dimension of those lists that guarantee at same time high GDLP cryptographic security and does not deter computational throughput, is the value *16*. Figure 1 displays the sixteen prime cycles, the defined $S_{381}$ group and the resulting 64-bits order of the cyclic subgroup <*p*>.

```
Dim= 16

Prime list= {2, 3, 5, 7, 11, 13, 17, 19, 23, 29, 31, 37, 41, 43, 47, 53}

Partition sum=
 {2, 5, 10, 17, 28, 41, 58, 77, 100, 129, 160, 197, 238, 281, 328, 381}

Primorial list=
 {2, 6, 30, 210, 2310, 30 030, 510 510, 9 699 690, 223 092 870,
  6 469 693 230, 200 560 490 130, 7 420 738 134 810, 304 250 263 527 210,
  13 082 761 331 670 030, 614 889 782 588 491 410, 32 589 158 477 190 044 730}
```

Figure 1. Parameter definitions. The last value of the second and third lists are respectively the selected order of the symmetric group and the order of the cyclic subgroup generated by a random permutation whose cycle lengths are given by the first list.

---

[1] Pedro Hecht: Maestría en Seguridad Informática, FCE-FCEyN-FI (Universidad Bs Aires) qubit101@gmail.com

## 3. Diffie-Hellman Protocol

Using above mentioned structures and operations, a generalized Diffie-Hellman key exchange is outlined at Figure 2.

(a) PUBLIC VALUES (preparation)
 $S_{381}$: permutation group (non-commutative); $|S_{381}| \equiv 381! \sim 3.596379714 \times 10^{819}$
 $p \in_R S_{381}$ generator of the <p> subgroup; $|<p>| \equiv \Omega = 32589158477190044730$
(b) PRIVATE VALUES
 ALICE_power $(a) \in_R \mathbb{Z}_\Omega$ ; ALICE private exponent
 BOB power $(b) \in_R \mathbb{Z}_\Omega$ ; BOB private exponent
(c) CALCULATED TOKENS interchanged
 ALICE_Token $(t_a) = p^a$
 BOB_Token $(t_b) = p^b$
(d) ALICE calculates the session key
 ALICE_key $(k) = (t_b)^a$
(e) BOB calculates the session key
 BOB_key $(k) = (t_a)^b$

Figure 2. Generalized Diffie-Hellman key exchange

The procedure is easy to follow with a numeric trial, as exposed separately in APPENDIX I, with same symbols as defined in Figure 2.

Using previous arguments and bearing in mind that neither polynomial time conventional DLP attack nor a quantum procedure against it is at hand; the computational security is assumed to be of 64-bits.

The protocol works fast, using a non-optimized Mathematica interpreted code implementing a "*square and multiply*" routine and working on a ®Core i5 PC @ 2.20GHz, the session mean time took 93,75 ms over a sample of 1000000 cycles.

## 4. ElGamal Cipher

Our version has his cryptographic security based on the double coset problem (DCP) or respectively, the decomposition problem (DP) as the one-way functions [7].

DCP or DP are supposedly hard challenges in group theory. As no quantum attack algorithm over symmetric groups is on sight and probably does not exist, these solutions do not belong to BQP complexity set. Of course, this statement should be proven; a challenge outside the purpose of present work.

We present here both approaches. The general procedure is outlined at following figures.

(a) PUBLIC VALUES (preparation)
 $S_{381}$: permutation group (non-commutative); $|S_{381}| \equiv 381! \sim 3.596379714 \times 10^{819}$
 $p \in_R S_{381}$ generator of the <p> subgroup; $|<p>| \equiv \Omega = 32589158477190044730$
 $g \in_R S_{381}$ auxiliar value
(b) PRIVATE VALUES
 $(m, n) \in_R (\mathbb{Z}_\Omega)^2$ ; $(p^m, p^n)$ ALICE private key
 $(r, s) \in_R (\mathbb{Z}_\Omega)^2$ ; $(p^r, p^s)$ BOB private key
(c) PUBLIC VALUES
 $p_A = p^m\, g\, p^n$
 $p_B = p^r\, g\, p^s$
(d) ALICE ciphers a message for BOB
 $t \in_R \mathbb{Z}_\Omega$ ; $k = p^t$ ALICE session key (secret)
 msg $\in \mathbb{Z}_\Omega$ ALICE selected message (converted factoradic number $< \Omega$)
 $(y_1, y_2)$ cipher of msg; $y_1 = k^m\, g\, k^n$ ; $y_2 = msg\,(k^m\, p_B\, k^n)$
(e) BOB deciphers the message
 $m = y_2\,(p^r\, y_1\, p^s)^{-1}$
  $= msg\,(k^m\, p_B\, k^n)(p^r\, y_1\, p^s)^{-1}$
  $= msg\,(k^m\,(p^r\, g\, p^s)\, k^n)(p^r\, y_1\, p^s)^{-1}$
  $= msg\,(p^r\,(k^m\, g\, k^n)\, p^s)(p^r\, y_1\, p^s)^{-1}$
  $= msg\,(p^r\, y_1\, p^s)(p^r\, y_1\, p^s)^{-1}$
  $= msg$

Figure 3a. Generalized ElGamal cipher using DCP as OWF

(a) PUBLIC VALUES (preparation)
 $S_{381}$: permutation group (non-commutative); $|S_{381}| \equiv 381! \sim 3.596379714 \times 10^{819}$
 $p \in_R S_{381}$ generator of the <p> subgroup; $|<p>| \equiv \Omega = 32589158477190044730$
 $q \in_R S_{381}$ generator of the <q> subgroup; $|<q>| \equiv \Omega = 32589158477190044730$
 $g \in_R S_{381}$ auxiliar value
(b) PRIVATE VALUES
 $(m, n) \in_R (\mathbb{Z}_\Omega)^2$ ; $(p^m, q^n)$ ALICE private key
 $(r, s) \in_R (\mathbb{Z}_\Omega)^2$ ; $(p^r, q^s)$ BOB private key
(c) PUBLIC VALUES
 $p_A = p^m\, g\, q^n$
 $p_B = p^r\, g\, q^s$
(d) ALICE ciphers a message for BOB
 $(t, u) \in_R (\mathbb{Z}_\Omega)^2$ ; $(k = p^t,\, l = q^u)$ ALICE session keys (secret)
 msg $\in \mathbb{Z}_\Omega$ ALICE selected message (converted factoradic number $< \Omega$)
 $(y_1, y_2)$ cipher of msg; $y_1 = k^m\, g\, l^n$ ; $y_2 = msg\,(k^m\, p_B\, l^n)$
(e) BOB deciphers the message
 $m = y_2\,(p^r\, y_1\, q^s)^{-1}$
  $= msg\,(k^m\, p_B\, l^n)(p^r\, y_1\, q^s)^{-1}$
  $= msg\,(k^m\,(p^r\, g\, q^s)\, l^n)(p^r\, y_1\, q^s)^{-1}$
  $= msg\,(p^r\,(k^m\, g\, l^n)\, q^s)(p^r\, y_1\, q^s)^{-1}$
  $= msg\,(p^r\, y_1\, q^s)(p^r\, y_1\, q^s)^{-1}$
  $= msg$

Figure 3b. Generalized ElGamal cipher using DP as OWF

Again, we proceed with a stepwise example. It is included at APPENDIX I using the DCP variation. All used symbols agreed with Figure 3a definitions.

## 5. Conclusions

We developed a PQC solution using the symmetric group as the embedding structure. This approach fits into non-commutative cryptography. The random selection of high order elements is easy to obtain and lead naturally into big cyclic subgroups, where the DCP or the DP are hard to solve. Permutation group operations like integer mapping, compositions (multiplications) and it powers, have easy solutions. It relies only on simple combinatorial operations, no need of arithmetic or big-number libraries. This feature would enable its use in low computational resources environments like cellphones, smart cards, etc.

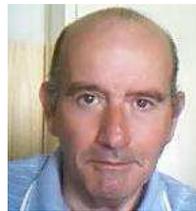
**Pedro Hecht** received an MSci in Information Technology at Escuela Superior de Investigación Operativa (ESIO-DIGID) and an PhD degree from Universidad Nacional de Buenos Aires (UBA). Currently, he is full professor of cryptography at Information Security Graduate School at UBA, EST (Army Engineering School) and IUPFA (Federal Police University), he is also a research fellow UBACyT and Director of EUDEBA editorial board of UBA. Current fields of interest are algebraic PQC solutions like non-commutative, non-associative structures.

# APPENDIX I: STEP BY STEP NUMERIC EXAMPLES OF THE GENERALIZED DIFFIE-HELLMAN KEY EXCHANGE AND GENERALIZED ELGAMAL CIPHER

**(1) Generalized Diffie-Hellman Key Exchange**

```
p= {3, 378, 273, 172, 96, 319, 341, 299, 229, 207, 147, 212,
193, 115, 73, 186, 338, 51, 45, 298, 255, 278, 159, 120,
234, 30, 178, 49, 166, 58, 345, 213, 28, 168, 141, 315,
264, 333, 85, 321, 199, 139, 21, 355, 258, 306, 380, 165,
373, 203, 287, 116, 289, 57, 324, 173, 217, 250, 191, 31,
313, 219, 366, 269, 295, 122, 190, 359, 260, 354, 251,
294, 358, 185, 241, 323, 145, 233, 8, 189, 314, 157, 90,
98, 99, 149, 4, 133, 224, 372, 195, 155, 18, 248, 257, 151,
29, 331, 222, 198, 230, 318, 134, 316, 348, 125, 92, 110,
184, 379, 161, 144, 361, 268, 206, 55, 72, 88, 169, 308,
288, 175, 87, 311, 352, 208, 216, 176, 26, 102, 86, 2, 104,
325, 296, 192, 187, 89, 215, 247, 182, 286, 275, 12, 156,
317, 94, 162, 62, 374, 283, 9, 280, 101, 267, 19, 231, 329,
103, 64, 126, 1, 59, 226, 150, 362, 47, 367, 84, 197, 106,
225, 16, 43, 282, 200, 279, 300, 261, 301, 109, 146, 75,
34, 277, 218, 181, 237, 270, 326, 40, 177, 284, 202, 246,
52, 97, 127, 17, 290, 68, 174, 108, 128, 220, 332, 105,
232, 253, 83, 276, 322, 160, 54, 60, 33, 91, 171, 204, 179,
140, 370, 153, 136, 292, 164, 375, 167, 310, 239, 293, 77,
262, 23, 15, 44, 65, 221, 80, 265, 281, 285, 74, 66, 291,
371, 312, 377, 263, 69, 50, 363, 71, 381, 266, 37, 78, 188,
163, 194, 309, 238, 138, 32, 61, 335, 112, 320, 107, 152,
337, 42, 259, 38, 100, 223, 346, 211, 79, 5, 376, 214, 249,
236, 254, 334, 70, 357, 342, 228, 274, 256, 36, 39, 10,
53, 113, 183, 22, 129, 353, 227, 252, 137, 119, 347, 356,
235, 11, 154, 121, 142, 132, 123, 330, 242, 25, 272, 148,
158, 6, 350, 118, 336, 340, 304, 48, 41, 364, 124, 114,
305, 180, 196, 170, 205, 351, 328, 209, 210, 27, 240, 243,
307, 56, 82, 368, 303, 245, 369, 81, 130, 365, 13, 67, 302,
76, 14, 201, 349, 131, 360, 143, 117, 7, 244, 46, 63, 35,
24, 339, 271, 344, 111, 327, 20, 135, 95, 93, 297, 343}
```

Figure 1. Random permutation $p$, generator of the cyclic subgroup $<p>$ belonging to $S_{381}$. This public value could be concerted in advance or transferred to the second entity by the initiator.

```
Cycle lengths=
{13, 41, 47, 23, 37, 11, 43, 53, 31, 19, 5, 7, 29, 17, 3, 2}

|<p>|= 32 589 158 477 190 044 730
```

Figure 2. Embedded cycle lengths of $p$ and cyclic subgroup $<p>$ order, both public and fixed parameters.

Once the generator is concerted, the protocol follows as usual with the selection of random secret exponents for each entity and subsequent exchange of public tokens.

```
ALICE_power= 16 967 309 044 902 469 564
BOB_power=   10 540 455 745 810 519 467
```

Figure 3. Alice and Bob randomly selected secret exponents $\{a,b\}$

```
ALICE_token= {162, 132, 1, 269, 279, 321, 43, 283, 154, 356, 309, 234,
306, 114, 84, 139, 199, 184, 216, 75, 38, 263, 337, 332, 372, 197,
255, 77, 341, 97, 102, 12, 232, 195, 325, 82, 112, 250, 73, 191,
328, 106, 274, 366, 33, 339, 380, 10, 145, 326, 34, 196, 296, 51,
116, 272, 287, 29, 163, 130, 265, 219, 251, 350, 344, 108, 66, 359,
362, 367, 67, 235, 98, 254, 376, 311, 100, 257, 151, 152, 32, 74,
172, 72, 358, 149, 64, 357, 276, 225, 354, 141, 109, 147, 378, 79,
7, 294, 14, 111, 80, 186, 81, 323, 227, 345, 35, 137, 217, 187, 303,
317, 361, 85, 268, 52, 308, 288, 364, 305, 315, 110, 160, 293, 56,
363, 126, 176, 170, 16, 86, 313, 76, 314, 377, 153, 214, 211, 125,
221, 340, 312, 150, 25, 198, 83, 11, 319, 62, 105, 8, 310, 192, 239,
182, 127, 185, 370, 351, 322, 252, 148, 259, 164, 207, 27, 47, 246,
117, 353, 31, 107, 42, 291, 379, 200, 5, 266, 220, 260, 57, 210,
281, 91, 381, 215, 54, 49, 9, 122, 59, 280, 347, 349, 13, 334, 365,
161, 41, 290, 68, 245, 304, 128, 336, 320, 302, 143, 44, 4, 89, 23,
212, 18, 352, 208, 70, 60, 204, 205, 238, 115, 136, 223, 92, 226,
237, 167, 101, 189, 277, 275, 78, 271, 169, 63, 373, 262, 270, 342,
20, 88, 285, 203, 26, 193, 140, 94, 22, 166, 190, 48, 355, 104, 333,
267, 95, 28, 273, 360, 179, 233, 278, 144, 240, 180, 146, 99, 369,
229, 256, 171, 3, 58, 374, 224, 343, 138, 96, 177, 298, 93, 299,
368, 133, 142, 168, 330, 53, 228, 30, 155, 346, 15, 307, 135, 113,
241, 249, 335, 194, 188, 327, 282, 329, 209, 45, 119, 261, 230,
36, 247, 61, 213, 157, 118, 90, 218, 6, 222, 40, 159, 121, 55, 123,
175, 295, 338, 24, 231, 39, 158, 69, 286, 301, 324, 37, 17, 236,
87, 21, 289, 242, 19, 318, 243, 253, 375, 129, 103, 264, 173, 202,
46, 244, 258, 124, 331, 201, 300, 131, 178, 165, 120, 174, 50, 371,
71, 134, 206, 284, 292, 156, 348, 65, 183, 248, 2, 181, 297, 316}
```

Figure 4. Alice public token $t_a = p^a$.

```
BOB_token= {259, 11, 163, 372, 283, 162, 38, 278, 101, 49, 312, 64, 379,
364, 320, 139, 328, 251, 275, 298, 69, 276, 369, 98, 322, 7, 180,
161, 255, 341, 102, 87, 111, 326, 267, 82, 314, 362, 332, 319, 17,
106, 250, 287, 100, 187, 149, 188, 126, 246, 50, 378, 55, 366, 257,
272, 244, 27, 321, 130, 179, 228, 34, 225, 33, 46, 354, 68, 300,
108, 91, 115, 158, 254, 241, 311, 252, 377, 22, 9, 340, 74, 234,
222, 305, 200, 90, 357, 192, 23, 122, 37, 71, 286, 94, 249, 21, 370,
119, 48, 189, 186, 141, 323, 344, 345, 256, 339, 67, 209, 10, 213,
204, 120, 117, 95, 14, 288, 85, 331, 315, 306, 83, 293, 56, 105,
150, 380, 365, 16, 176, 309, 76, 182, 52, 279, 236, 136, 125, 342,
112, 61, 375, 160, 363, 12, 142, 3, 290, 65, 263, 154, 96, 80, 264,
143, 185, 15, 35, 172, 207, 273, 40, 226, 237, 266, 86, 304, 99,
174, 31, 271, 42, 58, 347, 297, 8, 301, 221, 349, 355, 144, 75, 190,
381, 215, 44, 127, 229, 13, 6, 79, 93, 26, 175, 2, 43, 165, 338, 113,
201, 30, 371, 47, 262, 72, 373, 348, 57, 25, 153, 269, 4, 63, 352,
374, 66, 60, 167, 238, 289, 169, 5, 177, 337, 164, 19, 131, 230,
270, 277, 303, 135, 350, 268, 51, 216, 53, 152, 205, 281, 88, 285,
367, 97, 282, 240, 334, 89, 178, 195, 302, 217, 104, 260, 81, 248,
198, 191, 129, 140, 296, 224, 123, 220, 194, 32, 308, 292, 310, 103,
171, 59, 166, 327, 280, 343, 223, 299, 151, 376, 368, 138, 18, 133,
313, 203, 330, 324, 361, 29, 351, 346, 206, 28, 116, 219, 183, 211,
353, 245, 156, 356, 284, 39, 181, 77, 114, 247, 239, 36, 265, 261,
210, 157, 118, 212, 218, 1, 235, 148, 107, 121, 78, 146, 193, 258,
199, 73, 231, 24, 294, 360, 132, 202, 233, 134, 41, 54, 317, 333,
336, 242, 232, 318, 243, 109, 307, 197, 92, 325, 173, 291, 110,
70, 145, 124, 329, 359, 170, 128, 335, 227, 358, 274, 168, 137,
184, 155, 84, 214, 159, 208, 295, 45, 20, 196, 147, 253, 62, 316}
```

Figure 5. Bob public token $t_b = p^b$

Finally, both obtain a common session key because $<p>$ has cyclic structure and powers commute.

```
ALICE_key= {319, 196, 148, 264, 79, 191, 197, 249, 229, 374, 2, 141, 122,
308, 370, 42, 41, 368, 302, 75, 7, 138, 210, 39, 325, 349, 29, 65,
30, 245, 130, 92, 237, 251, 234, 285, 269, 43, 329, 163, 338, 171,
365, 54, 227, 108, 149, 275, 295, 34, 63, 135, 238, 236, 53, 318,
44, 291, 273, 352, 142, 228, 18, 112, 165, 70, 217, 68, 38, 244,
109, 320, 24, 76, 376, 346, 344, 324, 283, 189, 225, 133, 81, 206,
364, 200, 37, 231, 223, 314, 67, 212, 347, 95, 116, 8, 26, 332, 117,
19, 230, 16, 90, 36, 161, 31, 12, 367, 253, 46, 216, 369, 204, 119,
235, 296, 268, 157, 14, 85, 74, 354, 256, 343, 345, 28, 258, 380,
360, 173, 176, 334, 293, 372, 233, 280, 371, 276, 106, 261, 23, 147,
145, 35, 307, 103, 378, 321, 290, 198, 299, 9, 177, 101, 322, 356,
104, 98, 83, 267, 33, 6, 3, 164, 100, 58, 86, 50, 115, 300, 60, 32,
272, 353, 13, 297, 96, 166, 61, 255, 209, 159, 281, 71, 323, 139,
339, 327, 270, 91, 259, 5, 175, 180, 190, 377, 129, 45, 328, 113,
201, 301, 168, 47, 240, 15, 111, 49, 187, 351, 224, 182, 155, 284,
125, 188, 355, 215, 167, 265, 179, 72, 192, 153, 213, 226, 252,
131, 310, 239, 316, 373, 336, 340, 222, 214, 48, 220, 80, 247, 20,
330, 357, 287, 194, 326, 309, 257, 278, 274, 184, 232, 181, 311,
341, 64, 55, 375, 1, 333, 313, 205, 211, 107, 312, 27, 350, 169,
144, 152, 87, 218, 162, 174, 156, 136, 118, 89, 151, 279, 298, 193,
22, 282, 124, 94, 366, 185, 221, 361, 202, 160, 242, 158, 150, 262,
219, 241, 263, 362, 266, 303, 208, 246, 358, 110, 105, 99, 132, 154,
243, 11, 286, 292, 254, 315, 134, 186, 40, 84, 59, 146, 82, 289,
271, 195, 143, 17, 331, 381, 305, 73, 21, 248, 178, 342, 4, 199,
137, 337, 97, 140, 288, 207, 102, 88, 379, 126, 260, 317, 172, 56,
335, 66, 57, 348, 277, 120, 359, 69, 128, 250, 77, 114, 170, 51,
203, 93, 25, 294, 304, 123, 10, 127, 363, 183, 78, 52, 306, 62, 121}
```

Figure 6. Alice $key=(t_b)^a$

```
BOB_key= {319, 196, 148, 264, 79, 191, 197, 249, 229, 374, 2, 141, 122,
308, 370, 42, 41, 368, 302, 75, 7, 138, 210, 39, 325, 349, 29, 65,
30, 245, 130, 92, 237, 251, 234, 285, 269, 43, 329, 163, 338, 171,
365, 54, 227, 108, 149, 275, 295, 34, 63, 135, 238, 236, 53, 318,
44, 291, 273, 352, 142, 228, 18, 112, 165, 70, 217, 68, 38, 244,
109, 320, 24, 76, 376, 346, 344, 324, 283, 189, 225, 133, 81, 206,
364, 200, 37, 231, 223, 314, 67, 212, 347, 95, 116, 8, 26, 332, 117,
19, 230, 16, 90, 36, 161, 31, 12, 367, 253, 46, 216, 369, 204, 119,
235, 296, 268, 157, 14, 85, 74, 354, 256, 343, 345, 28, 258, 380,
360, 173, 176, 334, 293, 372, 233, 280, 371, 276, 106, 261, 23, 147,
145, 35, 307, 103, 378, 321, 290, 198, 299, 9, 177, 101, 322, 356,
104, 98, 83, 267, 33, 6, 3, 164, 100, 58, 86, 50, 115, 300, 60, 32,
272, 353, 13, 297, 96, 166, 61, 255, 209, 159, 281, 71, 323, 139,
339, 327, 270, 91, 259, 5, 175, 180, 190, 377, 129, 45, 328, 113,
201, 301, 168, 47, 240, 15, 111, 49, 187, 351, 224, 182, 155, 284,
125, 188, 355, 215, 167, 265, 179, 72, 192, 153, 213, 226, 252,
131, 310, 239, 316, 373, 336, 340, 222, 214, 48, 220, 80, 247, 20,
330, 357, 287, 194, 326, 309, 257, 278, 274, 184, 232, 181, 311,
341, 64, 55, 375, 1, 333, 313, 205, 211, 107, 312, 27, 350, 169,
144, 152, 87, 218, 162, 174, 156, 136, 118, 89, 151, 279, 298, 193,
22, 282, 124, 94, 366, 185, 221, 361, 202, 160, 242, 158, 150, 262,
219, 241, 263, 362, 266, 303, 208, 246, 358, 110, 105, 99, 132, 154,
243, 11, 286, 292, 254, 315, 134, 186, 40, 84, 59, 146, 82, 289,
271, 195, 143, 17, 331, 381, 305, 73, 21, 248, 178, 342, 4, 199,
137, 337, 97, 140, 288, 207, 102, 88, 379, 126, 260, 317, 172, 56,
335, 66, 57, 348, 277, 120, 359, 69, 128, 250, 77, 114, 170, 51,
203, 93, 25, 294, 304, 123, 10, 127, 363, 183, 78, 52, 306, 62, 121}
```

Figure 7. Bob $key=(t_a)^b$

(2) **Generalized ElGamal Cipher**
Here we use the Fig 3a. variation based on DCP.

```
p = {119, 14, 56, 261, 8, 337, 146, 301, 220, 257, 40, 296, 175,
127, 331, 345, 39, 333, 151, 193, 343, 25, 291, 86, 346,
28, 120, 340, 207, 65, 143, 232, 63, 162, 293, 235, 150,
321, 44, 292, 137, 189, 264, 376, 187, 107, 263, 164, 205,
266, 171, 269, 60, 105, 1, 88, 62, 96, 202, 366, 103, 262,
362, 255, 186, 306, 316, 368, 174, 4, 190, 280, 352, 233,
169, 341, 73, 364, 50, 160, 267, 330, 168, 59, 381, 83,
48, 116, 295, 348, 322, 20, 203, 26, 260, 329, 112, 15,
259, 300, 215, 155, 145, 372, 325, 181, 122, 130, 339, 84,
282, 276, 315, 375, 270, 192, 108, 228, 71, 159, 109, 82,
252, 312, 49, 177, 332, 284, 288, 89, 360, 297, 131, 173,
251, 19, 153, 361, 167, 247, 74, 179, 268, 290, 311, 45,
221, 123, 114, 320, 194, 226, 314, 373, 234, 10, 351, 13,
236, 57, 210, 342, 317, 61, 115, 237, 95, 11, 231, 283, 38,
318, 201, 7, 377, 197, 78, 121, 3, 46, 91, 240, 310, 363,
213, 225, 68, 239, 369, 55, 102, 309, 87, 279, 178, 93,
298, 195, 334, 99, 371, 176, 244, 277, 294, 191, 188, 286,
180, 113, 305, 357, 79, 66, 158, 354, 242, 129, 229, 141,
336, 94, 35, 198, 111, 72, 148, 253, 157, 36, 152, 58, 69,
344, 101, 245, 140, 350, 9, 182, 184, 142, 149, 70, 246,
230, 5, 43, 224, 138, 165, 2, 51, 90, 356, 374, 211, 76,
30, 216, 326, 274, 204, 313, 104, 307, 85, 328, 136, 128,
23, 227, 133, 97, 278, 24, 54, 75, 200, 254, 16, 80, 32,
208, 47, 92, 117, 265, 41, 308, 243, 118, 370, 81, 365, 17,
380, 6, 323, 359, 110, 281, 172, 37, 347, 12, 209, 64,
27, 135, 53, 353, 125, 219, 223, 335, 271, 139, 327, 147,
18, 156, 250, 166, 67, 256, 52, 124, 302, 98, 222, 199,
132, 338, 134, 34, 378, 170, 367, 29, 249, 272, 248, 355,
31, 144, 218, 319, 324, 154, 241, 206, 33, 349, 273, 42,
303, 77, 183, 258, 100, 299, 214, 217, 285, 21, 161, 185,
212, 22, 163, 379, 358, 106, 196, 287, 275, 238, 304, 289}

Cycle lengths=
{5, 17, 41, 31, 47, 37, 43, 53, 3, 29, 19, 13, 11, 7, 23, 2}

|<p>|= 32 589 158 477 190 044 730
```

Figure 8. Public $<p>$ generator.

```
g = {310, 333, 136, 214, 36, 27, 336, 308, 300, 53, 228, 298,
246, 5, 61, 205, 344, 305, 175, 288, 238, 369, 9, 164,
245, 151, 265, 125, 131, 100, 48, 155, 198, 17, 179, 75,
93, 118, 41, 81, 45, 269, 171, 76, 54, 154, 63, 234, 78,
351, 376, 352, 56, 181, 3, 231, 43, 349, 299, 301, 222,
46, 10, 176, 213, 263, 16, 343, 91, 367, 227, 83, 235,
57, 229, 354, 280, 158, 127, 98, 295, 193, 144, 242, 226,
243, 148, 129, 355, 203, 97, 327, 161, 88, 19, 79, 318,
277, 143, 8, 68, 162, 106, 262, 197, 254, 102, 12, 4, 30,
225, 87, 138, 200, 59, 220, 282, 374, 50, 291, 248, 297,
60, 58, 253, 140, 167, 69, 182, 360, 117, 145, 208, 323,
137, 123, 11, 232, 2, 24, 15, 84, 92, 99, 44, 335, 271,
316, 187, 377, 135, 375, 339, 180, 322, 107, 174, 34,
14, 289, 267, 320, 223, 114, 361, 77, 95, 266, 348, 317,
312, 204, 120, 304, 185, 237, 356, 279, 74, 353, 274,
156, 80, 270, 247, 324, 65, 85, 207, 96, 18, 257, 303,
307, 132, 328, 272, 7, 373, 134, 379, 217, 331, 230, 211,
251, 105, 89, 160, 366, 233, 313, 284, 290, 215, 330,
285, 130, 1, 29, 195, 112, 276, 236, 283, 31, 32, 337,
216, 371, 47, 302, 169, 190, 347, 115, 212, 94, 326, 86,
306, 170, 172, 186, 259, 255, 150, 258, 6, 196, 219, 192,
113, 28, 345, 51, 110, 342, 378, 72, 116, 321, 39, 218,
66, 157, 71, 52, 241, 273, 20, 221, 152, 202, 368, 142,
239, 249, 358, 104, 13, 188, 178, 126, 325, 309, 264,
199, 319, 311, 240, 108, 122, 372, 35, 119, 124, 73, 67,
191, 33, 149, 37, 363, 334, 42, 244, 23, 294, 184, 359,
338, 101, 26, 281, 332, 153, 292, 147, 341, 256, 286,
121, 350, 275, 139, 365, 194, 278, 25, 64, 340, 22, 189,
362, 183, 329, 364, 103, 370, 287, 268, 133, 55, 168,
141, 293, 381, 260, 49, 165, 21, 314, 111, 146, 163, 177,
357, 315, 252, 62, 209, 380, 166, 201, 296, 224, 70, 261,
109, 206, 40, 159, 173, 250, 90, 210, 38, 128, 82, 346}

Cycle lengths= {248, 93, 6, 29, 4}

|<g>| = 21 576
```

Figure 8. Public auxiliar permutation

```
ALICE private keys

m= 9 427 189 104 773 785 613   n= 26 477 403 901 985 527 977

p^m = {245, 17, 273, 205, 133, 351, 307, 129, 107, 63, 377, 379,
141, 343, 275, 41, 310, 320, 261, 237, 187, 55, 156, 153,
362, 175, 165, 178, 301, 94, 98, 246, 328, 1, 82, 208, 290,
319, 154, 304, 180, 269, 93, 100, 353, 79, 231, 277, 206,
31, 40, 302, 10, 314, 270, 53, 37, 298, 66, 250, 368, 96,
47, 74, 142, 62, 39, 364, 99, 278, 281, 347, 69, 303, 89, 8,
155, 251, 2, 324, 363, 109, 293, 195, 183, 159, 224, 284,
104, 295, 143, 58, 311, 213, 369, 139, 92, 350, 297, 88,
338, 114, 111, 286, 130, 225, 164, 201, 16, 238, 23, 367,
357, 103, 265, 123, 198, 147, 226, 288, 125, 292, 375, 12,
279, 170, 333, 144, 126, 342, 33, 110, 77, 254, 247, 152,
150, 177, 34, 317, 325, 221, 124, 122, 30, 321, 64, 236,
223, 131, 185, 374, 76, 127, 255, 24, 148, 219, 68, 90,
359, 200, 105, 339, 234, 32, 22, 112, 274, 145, 161, 5, 20,
316, 116, 57, 189, 191, 193, 167, 26, 140, 186, 378, 220,
204, 276, 336, 264, 326, 162, 352, 4, 86, 257, 291, 360,
194, 313, 106, 72, 135, 35, 36, 45, 216, 218, 280, 376,
282, 54, 171, 84, 211, 215, 146, 137, 258, 25, 60, 192,
70, 197, 6, 9, 80, 13, 210, 355, 75, 56, 356, 181, 239, 91,
27, 43, 65, 78, 373, 306, 132, 14, 172, 209, 15, 29, 253,
50, 120, 330, 52, 358, 283, 61, 372, 21, 113, 160, 235,
136, 309, 46, 248, 361, 87, 315, 232, 241, 38, 176, 252,
118, 151, 337, 272, 49, 119, 18, 322, 346, 7, 300, 242, 11,
168, 121, 212, 203, 233, 199, 365, 318, 157, 380, 85, 108,
128, 263, 102, 196, 163, 214, 332, 341, 42, 243, 182, 266,
259, 169, 267, 312, 19, 381, 115, 44, 327, 331, 134, 294,
345, 28, 230, 228, 158, 349, 240, 296, 335, 179, 59, 344,
287, 256, 244, 366, 71, 354, 299, 323, 138, 166, 184, 222,
285, 97, 202, 73, 371, 48, 149, 95, 308, 262, 207, 334,
305, 348, 268, 190, 51, 340, 227, 260, 188, 249, 329, 3,
67, 229, 174, 101, 271, 173, 81, 117, 289, 83, 370, 217}

p^n = {267, 101, 77, 188, 370, 152, 376, 18, 189, 231, 219, 99,
96, 340, 2, 7, 338, 197, 36, 112, 156, 343, 334, 121, 359,
230, 375, 302, 146, 300, 147, 46, 256, 312, 228, 109, 119,
246, 194, 135, 307, 269, 249, 178, 244, 240, 53, 271, 57,
118, 202, 129, 47, 186, 166, 63, 278, 91, 301, 262, 333,
291, 56, 368, 164, 196, 198, 59, 58, 55, 361, 292, 92, 329,
4, 279, 39, 43, 373, 331, 137, 210, 108, 103, 289, 304, 261,
162, 205, 285, 349, 235, 50, 102, 5, 199, 260, 64, 298,
191, 115, 218, 113, 45, 145, 138, 264, 144, 282, 9, 357,
369, 65, 258, 60, 89, 362, 255, 38, 266, 352, 379, 104,
69, 149, 134, 159, 124, 320, 30, 335, 225, 67, 268, 206,
280, 251, 213, 313, 323, 139, 339, 73, 12, 283, 233, 61,
116, 140, 330, 371, 322, 125, 86, 154, 287, 175, 171, 332,
346, 351, 130, 170, 248, 173, 263, 14, 95, 98, 254, 6, 380,
168, 185, 75, 70, 84, 163, 337, 243, 324, 341, 203, 227,
229, 35, 24, 160, 195, 288, 105, 308, 366, 40, 111, 294,
44, 51, 148, 342, 122, 49, 325, 82, 336, 37, 132, 16, 381,
245, 183, 224, 114, 85, 215, 290, 78, 110, 161, 355, 354,
22, 317, 136, 177, 319, 62, 1, 214, 193, 10, 284, 80, 20,
327, 123, 367, 107, 237, 174, 306, 106, 363, 295, 315, 79,
216, 21, 273, 309, 93, 8, 187, 232, 127, 274, 23, 238, 281,
318, 208, 54, 328, 257, 250, 19, 165, 88, 241, 32, 222, 76,
155, 350, 17, 153, 176, 270, 223, 41, 190, 209, 356, 200,
158, 353, 192, 87, 141, 226, 157, 83, 201, 26, 11, 378,
128, 143, 33, 207, 321, 126, 296, 247, 272, 42, 81, 305,
314, 71, 31, 27, 236, 204, 234, 220, 28, 72, 15, 360, 181,
180, 52, 179, 34, 212, 347, 372, 184, 311, 275, 29, 68,
221, 169, 90, 310, 265, 253, 131, 252, 94, 299, 364, 167,
326, 365, 277, 97, 74, 374, 182, 172, 358, 211, 242, 142,
276, 48, 100, 120, 348, 150, 66, 293, 259, 3, 344, 133,
117, 303, 151, 316, 345, 286, 217, 25, 13, 297, 377, 239}
```

Figure 9. Alice private values

```
BOB private keys

r= 14 090 847 924 998 838 332    s= 22 570 145 711 539 886 927

pʳ = {245, 17, 273, 205, 133, 351, 307, 129, 107, 63, 377, 379,
141, 343, 275, 41, 310, 320, 261, 237, 187, 55, 156, 153,
362, 175, 165, 178, 301, 94, 98, 246, 328, 1, 82, 208, 290,
319, 154, 304, 180, 269, 93, 100, 353, 79, 231, 277, 206,
31, 40, 302, 10, 314, 270, 53, 37, 298, 66, 250, 368, 96,
47, 74, 142, 62, 39, 364, 99, 278, 281, 347, 69, 303, 89, 8,
155, 251, 2, 324, 363, 109, 293, 195, 183, 159, 224, 284,
104, 295, 143, 58, 311, 213, 369, 139, 92, 350, 297, 88,
338, 114, 111, 286, 130, 225, 164, 201, 16, 238, 23, 367,
357, 103, 265, 123, 198, 147, 226, 288, 125, 292, 375, 12,
279, 170, 333, 144, 126, 342, 33, 110, 77, 254, 247, 152,
150, 177, 34, 317, 325, 221, 124, 122, 30, 321, 64, 236,
223, 131, 185, 374, 76, 127, 255, 24, 148, 219, 68, 90,
359, 200, 105, 339, 234, 32, 22, 112, 274, 145, 161, 5, 20,
316, 116, 57, 189, 191, 193, 167, 26, 140, 186, 378, 220,
204, 276, 336, 264, 326, 162, 352, 4, 86, 257, 291, 360,
194, 313, 106, 72, 135, 35, 36, 45, 216, 218, 280, 376,
282, 54, 171, 84, 211, 215, 146, 137, 258, 25, 60, 192,
70, 197, 6, 9, 80, 13, 210, 355, 75, 56, 356, 181, 239, 91,
27, 43, 65, 78, 373, 306, 132, 14, 172, 209, 15, 29, 253,
50, 120, 330, 52, 358, 283, 61, 372, 21, 113, 160, 235,
136, 309, 46, 248, 361, 87, 315, 232, 241, 38, 176, 252,
118, 151, 337, 272, 49, 119, 18, 322, 346, 7, 300, 242, 11,
168, 121, 212, 203, 233, 199, 365, 318, 157, 380, 85, 108,
128, 263, 102, 196, 163, 214, 332, 341, 42, 243, 182, 266,
259, 169, 267, 312, 19, 381, 115, 44, 327, 331, 134, 294,
345, 28, 230, 228, 158, 349, 240, 296, 335, 179, 59, 344,
287, 256, 244, 366, 71, 354, 299, 323, 138, 166, 184, 222,
285, 97, 202, 73, 371, 48, 149, 95, 308, 262, 207, 334,
305, 348, 268, 190, 51, 340, 227, 260, 188, 249, 329, 3,
67, 229, 174, 101, 271, 173, 81, 117, 289, 83, 370, 217}

pˢ = {267, 101, 77, 188, 370, 152, 376, 18, 189, 231, 219, 99,
96, 340, 2, 7, 338, 197, 36, 112, 156, 343, 334, 121, 359,
230, 375, 302, 146, 300, 147, 46, 256, 312, 228, 109, 119,
246, 194, 135, 307, 269, 249, 178, 244, 240, 53, 271, 57,
118, 202, 129, 47, 186, 166, 63, 278, 91, 301, 262, 333,
291, 56, 368, 164, 196, 198, 59, 58, 55, 361, 292, 92, 329,
4, 279, 39, 43, 373, 331, 137, 210, 108, 103, 289, 304, 261,
162, 205, 285, 349, 235, 50, 102, 5, 199, 260, 64, 298,
191, 115, 218, 113, 45, 145, 138, 264, 144, 282, 9, 357,
369, 65, 258, 60, 89, 362, 255, 38, 266, 352, 379, 104,
69, 149, 134, 159, 124, 320, 30, 335, 225, 67, 268, 206,
280, 251, 213, 313, 323, 139, 339, 73, 12, 283, 233, 61,
116, 140, 330, 371, 322, 125, 86, 154, 287, 175, 171, 332,
346, 351, 130, 170, 248, 173, 263, 14, 95, 98, 254, 6, 380,
168, 185, 75, 70, 84, 163, 337, 243, 324, 341, 203, 227,
229, 35, 24, 160, 195, 288, 105, 308, 366, 40, 111, 294,
44, 51, 148, 342, 122, 49, 325, 82, 336, 37, 132, 16, 381,
245, 183, 224, 114, 85, 215, 290, 78, 110, 161, 355, 354,
22, 317, 136, 177, 319, 62, 1, 214, 193, 10, 284, 80, 20,
327, 123, 367, 107, 237, 174, 306, 106, 363, 295, 315, 79,
216, 21, 273, 309, 93, 8, 187, 232, 127, 274, 23, 238, 281,
318, 208, 54, 328, 257, 250, 19, 165, 88, 241, 32, 222, 76,
155, 350, 17, 153, 176, 270, 223, 41, 190, 209, 356, 200,
158, 353, 192, 87, 141, 226, 157, 83, 201, 26, 11, 378,
128, 143, 33, 207, 321, 126, 296, 247, 272, 42, 81, 305,
314, 71, 31, 27, 236, 204, 234, 220, 28, 72, 15, 360, 181,
180, 52, 179, 34, 212, 347, 372, 184, 311, 275, 29, 68,
221, 169, 90, 310, 265, 253, 131, 252, 94, 299, 364, 167,
326, 365, 277, 97, 74, 374, 182, 172, 358, 211, 242, 142,
276, 48, 100, 120, 348, 150, 66, 293, 259, 3, 344, 133,
117, 303, 151, 316, 345, 286, 217, 25, 13, 297, 377, 239}
```

Figure 10. Bob private values

```
p_A = {222, 249, 90, 235, 5, 43, 127, 56, 348, 307, 331, 160, 169,
79, 61, 151, 15, 18, 179, 78, 97, 8, 121, 74, 361, 156, 32,
252, 209, 92, 109, 271, 344, 363, 68, 113, 24, 36, 40, 294,
214, 207, 255, 2, 125, 186, 230, 334, 288, 81, 150, 197,
309, 332, 244, 167, 4, 280, 353, 216, 304, 380, 258, 196,
57, 108, 102, 69, 124, 134, 101, 66, 91, 279, 321, 49, 270,
58, 313, 131, 379, 137, 292, 276, 34, 342, 377, 350, 275,
59, 99, 157, 323, 263, 176, 46, 72, 329, 315, 64, 153, 119,
94, 371, 107, 245, 358, 253, 144, 106, 82, 308, 351, 360,
180, 356, 28, 63, 195, 162, 53, 133, 89, 173, 1, 60, 170,
76, 232, 330, 365, 210, 261, 268, 272, 16, 319, 183, 359,
199, 322, 117, 143, 155, 264, 42, 374, 220, 370, 233, 9,
85, 75, 41, 71, 375, 128, 241, 105, 135, 190, 203, 123,
378, 223, 178, 118, 368, 346, 192, 87, 239, 314, 11, 129,
103, 260, 326, 30, 193, 52, 201, 266, 362, 247, 112, 283,
98, 29, 242, 291, 301, 159, 234, 277, 13, 231, 215, 31,
88, 27, 189, 352, 246, 225, 110, 187, 325, 181, 211, 311,
284, 55, 285, 237, 354, 115, 317, 33, 23, 95, 149, 62, 226,
122, 290, 202, 19, 3, 338, 50, 140, 152, 7, 349, 219, 67,
224, 171, 345, 6, 83, 302, 343, 262, 218, 80, 335, 198,
305, 318, 194, 166, 324, 138, 221, 86, 312, 48, 355, 100,
200, 281, 381, 111, 146, 25, 132, 217, 267, 145, 12, 228,
282, 114, 93, 366, 126, 376, 136, 257, 26, 161, 174, 168,
130, 310, 172, 328, 188, 300, 206, 369, 299, 256, 250,
182, 298, 54, 14, 154, 373, 65, 238, 327, 273, 163, 248,
116, 259, 303, 333, 251, 236, 165, 364, 212, 44, 265, 372,
164, 17, 229, 84, 142, 337, 147, 320, 296, 22, 287, 175,
96, 20, 205, 70, 77, 120, 316, 336, 185, 254, 289, 269,
340, 35, 51, 341, 274, 38, 227, 213, 243, 204, 10, 191, 21,
139, 177, 184, 295, 104, 339, 39, 367, 347, 158, 73, 357,
47, 297, 278, 141, 286, 45, 306, 208, 240, 293, 37, 148}

p_B = {148, 3, 230, 16, 266, 159, 255, 335, 188, 245, 350, 155,
13, 297, 170, 264, 233, 92, 268, 141, 5, 269, 99, 267, 279,
142, 161, 194, 183, 146, 300, 130, 349, 333, 59, 204, 171,
352, 226, 93, 281, 67, 132, 52, 381, 210, 219, 113, 51, 4,
301, 296, 358, 326, 307, 166, 270, 227, 60, 328, 287, 295,
221, 80, 331, 58, 24, 354, 139, 179, 217, 274, 282, 262,
289, 48, 180, 329, 361, 263, 129, 259, 318, 242, 55, 175,
309, 120, 164, 314, 341, 250, 225, 276, 235, 362, 10, 88,
197, 355, 288, 38, 56, 138, 286, 323, 304, 375, 167, 372,
144, 12, 378, 112, 91, 79, 348, 228, 34, 19, 211, 212, 35,
162, 28, 109, 312, 18, 321, 127, 371, 244, 370, 27, 23,
173, 157, 71, 346, 46, 6, 156, 252, 223, 303, 32, 29, 298,
97, 201, 163, 380, 40, 186, 280, 231, 104, 311, 275, 243,
98, 277, 121, 364, 102, 110, 181, 325, 377, 89, 273, 240,
368, 158, 2, 232, 369, 236, 154, 366, 25, 83, 176, 305,
332, 41, 316, 26, 124, 133, 74, 238, 47, 216, 374, 15, 95,
86, 209, 239, 128, 39, 21, 200, 66, 119, 103, 33, 224, 62,
78, 136, 258, 36, 278, 11, 22, 306, 327, 203, 313, 367,
308, 42, 344, 177, 336, 172, 134, 81, 182, 96, 140, 246,
82, 254, 191, 85, 205, 241, 184, 338, 152, 126, 68, 363,
302, 87, 63, 107, 123, 198, 149, 293, 17, 339, 343, 315,
111, 294, 122, 284, 37, 237, 247, 54, 117, 101, 271, 94,
351, 64, 330, 106, 340, 253, 234, 70, 50, 292, 57, 147, 73,
196, 310, 153, 359, 373, 31, 84, 9, 249, 43, 299, 30, 137,
215, 376, 353, 151, 337, 114, 187, 108, 61, 257, 190, 290,
248, 285, 90, 44, 334, 202, 199, 365, 322, 291, 115, 189,
360, 185, 324, 220, 265, 222, 256, 193, 131, 20, 208, 345,
342, 77, 143, 213, 45, 168, 65, 49, 347, 317, 207, 125,
76, 100, 320, 260, 150, 195, 283, 8, 1, 229, 206, 356, 7,
165, 169, 178, 319, 14, 53, 192, 116, 105, 357, 174, 145,
214, 251, 72, 272, 118, 135, 379, 160, 218, 261, 69, 75}
```

Figure 11. Public keys (Alice, Bob)

```
t= 9 700 531 854 857 717 671

k= {119, 28, 101, 223, 140, 197, 219, 247, 187, 319, 30, 376,
133, 340, 331, 184, 93, 106, 38, 164, 2, 83, 303, 279,
168, 227, 142, 148, 174, 201, 40, 170, 290, 6, 42, 309,
132, 57, 203, 65, 49, 79, 102, 244, 157, 100, 366, 145,
85, 17, 80, 253, 124, 204, 1, 215, 328, 32, 110, 312, 135,
52, 308, 287, 371, 66, 370, 154, 113, 316, 190, 254, 82,
161, 231, 233, 122, 78, 296, 143, 41, 77, 99, 301, 153,
200, 103, 158, 159, 218, 114, 48, 261, 272, 211, 232,
19, 15, 134, 180, 255, 264, 251, 163, 277, 243, 300, 27,
178, 8, 346, 151, 350, 23, 208, 13, 310, 225, 71, 179,
195, 359, 94, 96, 267, 126, 367, 92, 265, 120, 295, 18,
89, 112, 270, 171, 205, 220, 10, 166, 368, 36, 292, 216,
165, 229, 262, 222, 271, 297, 321, 72, 294, 76, 313, 349,
361, 273, 3, 268, 379, 337, 335, 311, 284, 105, 257, 259,
152, 109, 160, 256, 276, 315, 131, 59, 177, 198, 235,
362, 149, 240, 217, 281, 307, 22, 351, 146, 50, 55, 43,
175, 61, 147, 224, 4, 202, 249, 91, 336, 24, 84, 326,
47, 381, 248, 7, 193, 63, 238, 167, 185, 12, 214, 356,
347, 246, 288, 250, 68, 274, 343, 357, 342, 25, 280, 332,
111, 138, 192, 226, 283, 210, 353, 64, 56, 325, 207, 45,
182, 302, 230, 318, 67, 88, 116, 237, 191, 162, 9, 128,
26, 282, 348, 117, 37, 139, 74, 173, 305, 35, 269, 285,
155, 372, 39, 137, 118, 51, 286, 358, 127, 130, 136, 75,
194, 263, 169, 221, 90, 241, 31, 339, 20, 60, 87, 183,
104, 125, 354, 172, 186, 189, 289, 236, 196, 333, 176,
345, 46, 5, 352, 327, 334, 95, 306, 44, 33, 377, 242,
115, 329, 234, 81, 324, 212, 338, 374, 156, 380, 62, 375,
355, 239, 54, 293, 304, 228, 58, 73, 98, 21, 181, 322,
378, 97, 298, 275, 121, 123, 69, 34, 14, 206, 363, 11,
260, 129, 323, 188, 258, 330, 344, 144, 299, 108, 213,
320, 107, 365, 141, 209, 16, 364, 245, 53, 252, 373, 266,
369, 86, 317, 341, 150, 291, 70, 360, 278, 29, 199, 314}

msg= {2, 302, 138, 297, 187, 73, 242, 37, 268, 152, 258, 59,
359, 180, 303, 237, 341, 190, 47, 244, 217, 10, 344, 68,
3, 334, 50, 360, 372, 48, 162, 98, 112, 28, 273, 251, 259,
375, 117, 71, 314, 61, 228, 332, 157, 193, 134, 274, 96,
60, 201, 195, 300, 223, 158, 84, 234, 254, 284, 374, 87,
145, 45, 29, 119, 204, 343, 276, 101, 22, 104, 264, 151,
227, 301, 129, 339, 143, 109, 324, 65, 235, 130, 323, 287,
82, 206, 315, 20, 319, 127, 325, 63, 125, 111, 95, 85, 296,
103, 131, 185, 14, 46, 140, 249, 289, 355, 304, 306, 317,
93, 252, 52, 133, 56, 349, 295, 221, 110, 8, 327, 239, 211,
350, 358, 175, 43, 214, 290, 337, 288, 12, 321, 189, 365,
328, 170, 226, 311, 354, 86, 270, 207, 173, 186, 160, 248,
142, 199, 76, 292, 356, 115, 74, 231, 194, 367, 320, 121,
126, 261, 58, 139, 40, 67, 150, 114, 210, 106, 305, 159, 83,
178, 240, 80, 148, 362, 318, 278, 70, 164, 163, 205, 219,
174, 78, 9, 262, 338, 123, 100, 38, 247, 4, 283, 377, 24,
149, 200, 279, 49, 313, 212, 97, 179, 161, 34, 168, 222,
116, 15, 277, 335, 177, 312, 16, 316, 18, 6, 137, 69, 238,
198, 64, 281, 220, 191, 7, 36, 89, 147, 122, 141, 370, 79,
183, 253, 269, 353, 154, 5, 307, 371, 25, 345, 308, 208,
379, 153, 373, 169, 197, 236, 132, 203, 340, 51, 92, 309,
128, 272, 215, 13, 266, 348, 209, 182, 124, 57, 245, 99,
88, 23, 213, 286, 329, 310, 120, 368, 233, 293, 331, 165,
72, 282, 250, 32, 333, 144, 265, 105, 364, 166, 113, 202,
146, 66, 342, 75, 118, 77, 26, 33, 246, 53, 271, 172, 224,
260, 378, 11, 196, 255, 102, 381, 291, 357, 181, 257, 62,
285, 41, 184, 366, 230, 322, 294, 298, 263, 232, 376, 380,
19, 351, 216, 135, 299, 347, 346, 81, 42, 241, 192, 44,
94, 256, 108, 229, 326, 176, 90, 91, 369, 243, 267, 30,
35, 156, 352, 107, 361, 275, 21, 54, 171, 330, 225, 218,
1, 167, 363, 336, 188, 17, 39, 280, 155, 27, 31, 136, 55}
```

Figure 11. Alice session key and message

```
y₁= {236, 52, 280, 143, 18, 319, 178, 158, 67, 140, 334, 214,
69, 294, 133, 374, 209, 83, 219, 172, 356, 123, 260, 174,
108, 171, 14, 110, 165, 363, 93, 237, 10, 176, 68, 141,
231, 296, 305, 150, 5, 344, 205, 332, 357, 225, 121, 125,
218, 98, 210, 372, 104, 250, 192, 361, 77, 257, 56, 309,
351, 196, 30, 129, 253, 249, 283, 238, 89, 177, 227, 28, 7,
277, 342, 184, 191, 287, 233, 193, 278, 102, 298, 279, 308,
326, 355, 362, 221, 217, 322, 222, 323, 310, 2, 290, 365,
87, 153, 63, 263, 119, 124, 369, 318, 147, 353, 144, 248,
330, 92, 132, 271, 179, 199, 85, 120, 159, 376, 135, 136,
368, 62, 317, 100, 170, 79, 380, 66, 185, 186, 304, 252,
31, 126, 39, 26, 122, 107, 50, 282, 268, 297, 114, 183, 41,
258, 340, 379, 160, 301, 112, 40, 343, 154, 42, 195, 17,
194, 256, 111, 70, 281, 146, 273, 163, 51, 168, 303, 48,
206, 164, 269, 27, 264, 3, 360, 375, 378, 320, 266, 97, 90,
21, 152, 312, 345, 81, 314, 202, 1, 53, 261, 259, 188, 116,
339, 36, 138, 15, 139, 216, 247, 4, 142, 23, 94, 234, 61,
338, 6, 366, 346, 73, 101, 328, 329, 149, 38, 84, 20, 113,
127, 285, 37, 348, 180, 354, 265, 359, 371, 155, 82, 118,
377, 131, 349, 34, 315, 33, 13, 76, 130, 313, 336, 335,
274, 175, 25, 71, 242, 64, 246, 137, 275, 300, 299, 311,
43, 46, 350, 11, 302, 145, 86, 99, 201, 321, 16, 254, 156,
32, 295, 381, 45, 47, 215, 226, 324, 223, 173, 88, 288,
148, 244, 358, 203, 80, 24, 213, 316, 166, 187, 55, 162,
255, 220, 272, 251, 241, 352, 286, 12, 229, 96, 307, 367,
151, 347, 327, 29, 59, 291, 78, 91, 115, 364, 197, 239, 57,
235, 224, 181, 211, 60, 212, 128, 22, 44, 106, 198, 105,
208, 240, 228, 284, 245, 74, 370, 189, 230, 333, 243, 72,
117, 103, 341, 182, 276, 157, 200, 169, 204, 75, 267, 270,
190, 161, 232, 95, 207, 325, 306, 109, 8, 58, 337, 331,
373, 35, 167, 49, 293, 134, 292, 262, 65, 54, 289, 19, 9}

y₂= {142, 315, 308, 61, 51, 322, 274, 377, 75, 6, 8, 53, 281,
57, 245, 156, 238, 294, 102, 260, 230, 248, 171, 17, 3, 321,
139, 375, 277, 270, 239, 257, 68, 268, 167, 132, 243, 207,
213, 269, 82, 42, 104, 278, 182, 249, 183, 62, 16, 217,
67, 48, 272, 370, 63, 332, 259, 313, 84, 66, 14, 327, 157,
107, 56, 284, 380, 198, 352, 262, 138, 335, 347, 242, 163,
103, 195, 148, 309, 144, 110, 127, 205, 18, 363, 254, 356,
52, 203, 271, 137, 20, 228, 176, 49, 191, 223, 37, 340,
159, 87, 113, 99, 354, 35, 210, 28, 247, 131, 134, 212,
258, 222, 351, 36, 379, 43, 117, 13, 147, 251, 101, 225,
50, 204, 208, 65, 241, 165, 368, 300, 286, 253, 80, 343,
296, 196, 285, 190, 100, 224, 312, 280, 206, 125, 334, 237,
108, 59, 178, 44, 126, 360, 235, 325, 369, 170, 302, 86,
304, 324, 371, 10, 211, 96, 319, 365, 311, 164, 279, 186,
273, 250, 121, 267, 297, 146, 12, 306, 151, 11, 114, 141,
122, 116, 83, 366, 150, 120, 378, 185, 78, 136, 339, 85,
350, 323, 172, 292, 153, 25, 299, 177, 41, 233, 106, 314,
240, 263, 305, 79, 231, 376, 320, 295, 359, 112, 261, 288,
158, 94, 98, 90, 26, 45, 130, 342, 345, 317, 194, 362, 316,
244, 39, 357, 128, 95, 331, 361, 252, 119, 218, 69, 124,
55, 64, 21, 133, 333, 200, 232, 215, 227, 355, 152, 72,
27, 77, 169, 181, 24, 381, 160, 123, 301, 175, 216, 373,
353, 92, 372, 289, 135, 291, 115, 246, 47, 337, 201, 179,
168, 71, 307, 23, 318, 54, 33, 367, 9, 19, 221, 255, 283,
202, 189, 184, 234, 193, 118, 180, 348, 4, 326, 46, 192,
275, 349, 1, 22, 298, 105, 97, 129, 364, 32, 74, 329, 155,
161, 31, 328, 154, 73, 199, 214, 40, 341, 374, 293, 60, 29,
209, 276, 226, 264, 197, 89, 81, 70, 330, 88, 149, 38, 346,
93, 290, 143, 236, 188, 336, 338, 219, 7, 287, 266, 109,
358, 5, 173, 58, 229, 265, 111, 30, 140, 256, 174, 282,
162, 2, 76, 344, 34, 310, 187, 91, 220, 166, 303, 145, 15}
```

Figure 12. ElGamal cipher pair (y₁,y₂)

```
msg = y₂ (pʳ y₁ pˢ)⁻¹ =
{2, 302, 138, 297, 187, 73, 242, 37, 268, 152, 258, 59, 359,
 180, 303, 237, 341, 190, 47, 244, 217, 10, 344, 68, 3,
 334, 50, 360, 372, 48, 162, 98, 112, 28, 273, 251, 259,
 375, 117, 71, 314, 61, 228, 332, 157, 193, 134, 274, 96,
 60, 201, 195, 300, 223, 158, 84, 234, 254, 284, 374, 87,
 145, 45, 29, 119, 204, 343, 276, 101, 22, 104, 264, 151,
 227, 301, 129, 339, 143, 109, 324, 65, 235, 130, 323,
 287, 82, 206, 315, 20, 319, 127, 325, 63, 125, 111, 95,
 85, 296, 103, 131, 185, 14, 46, 140, 249, 289, 355, 304,
 306, 317, 93, 252, 52, 133, 56, 349, 295, 221, 110, 8,
 327, 239, 211, 350, 358, 175, 43, 214, 290, 337, 288, 12,
 321, 189, 365, 328, 170, 226, 311, 354, 86, 270, 207,
 173, 186, 160, 248, 142, 199, 76, 292, 356, 115, 74, 231,
 194, 367, 320, 121, 126, 261, 58, 139, 40, 67, 150, 114,
 210, 106, 305, 159, 83, 178, 240, 80, 148, 362, 318, 278,
 70, 164, 163, 205, 219, 174, 78, 9, 262, 338, 123, 100,
 38, 247, 4, 283, 377, 24, 149, 200, 279, 49, 313, 212,
 97, 179, 161, 34, 168, 222, 116, 15, 277, 335, 177, 312,
 16, 316, 18, 6, 137, 69, 238, 198, 64, 281, 220, 191, 7,
 36, 89, 147, 122, 141, 370, 79, 183, 253, 269, 353, 154,
 5, 307, 371, 25, 345, 308, 208, 379, 153, 373, 169, 197,
 236, 132, 203, 340, 51, 92, 309, 128, 272, 215, 13, 266,
 348, 209, 182, 124, 57, 245, 99, 88, 23, 213, 286, 329,
 310, 120, 368, 233, 293, 331, 165, 72, 282, 250, 32,
 333, 144, 265, 105, 364, 166, 113, 202, 146, 66, 342,
 75, 118, 77, 26, 33, 246, 53, 271, 172, 224, 260, 378,
 11, 196, 255, 102, 381, 291, 357, 181, 257, 62, 285, 41,
 184, 366, 230, 322, 294, 298, 263, 232, 376, 380, 19,
 351, 216, 135, 299, 347, 346, 81, 42, 241, 192, 44, 94,
 256, 108, 229, 326, 176, 90, 91, 369, 243, 267, 30, 35,
 156, 352, 107, 361, 275, 21, 54, 171, 330, 225, 218, 1,
 167, 363, 336, 188, 17, 39, 280, 155, 27, 31, 136, 55}
```

Figure 13. Bob recovered message